\documentclass[12pt,epsf,amssymb,qsymbols]{article}
\usepackage{tabularx}
\usepackage{array}
\usepackage{graphics}
\usepackage{graphicx}
\usepackage{epsfig}
\usepackage{amsmath}
\usepackage{amssymb}
\makeatletter

\usepackage{verbatim}

\newcommand{\msbar}{{\overline{\rm MS}}}
\newcommand{\ri}{{\rm RI/MOM}}
\newcommand{\bea}{\begin{eqnarray}}
\newcommand{\eea}{\end{eqnarray}}
\newcommand{\beq}{\begin{equation}}
\newcommand{\eeq}{\end{equation}}
\newcommand{\gev}{{\rm GeV}}

\newcommand{\pdir}{p\kern -5.2pt\raise 0.2ex\hbox {/}}
\newcommand{\vdir}{v\kern -5.75pt\raise 0.15ex\hbox {/}}
\newcommand{\kdir}{k\kern -5.75pt\raise 0.15ex\hbox {/}}
\newcommand{\epsdir}{\epsilon\kern -5.0pt\raise 0.15ex\hbox {/}}
\newcommand{\bvdir}{\bar{v}\kern -5.75pt\raise 0.15ex\hbox {/}}
\newcommand{\Ddir}{D\kern -7.75pt\raise 0.20ex\hbox {/}}
\newcommand{\ldir}{l\kern -5.0pt\raise 0.2ex\hbox{/}}
\newcommand{\varepsdir}{\varepsilon\kern -5.5pt\raise 0.15ex\hbox{/}}

\makeatother

\begin{document}
\thispagestyle{empty} 
\begin{flushright}
\begin{tabular}{l}
{\tt RM3-th/01-9}\\
{\tt Roma-1317/01}
\end{tabular}
\end{flushright}
\begin{center}
\vskip 1.2cm\par
{\par\centering \LARGE \bf Charm quark mass}\\
\vskip 0.75cm\par
{\par\centering \large  
\sc D.~Be\'cirevi\'c~$^a$, V.~Lubicz~$^b$ and G.~Martinelli~$^a$}
{\par\centering \vskip 0.5 cm\par}
{\sl 
$^a$ Dip. di Fisica, Univ. di Roma ``La Sapienza" and INFN,
Sezione di Roma,\\
Piazzale Aldo Moro 2, I-00185 Rome, Italy. \\                                   
\vspace{.25cm}
$^b$ Dip. di Fisica, Univ. di Roma Tre and INFN,
Sezione di Roma III, \\
Via della Vasca Navale 84, I-00146 Rome, Italy.}\\
 
{\vskip 0.25cm\par}
\end{center}

\begin{abstract}
We report on the result for the charm quark mass as obtained from our lattice 
QCD computation in the quenched approximation. Our result is 
$m_c^\msbar(m_c)=1.26(4)(12)$~GeV.
\end{abstract}
\vskip 2.2 cm 
\setcounter{page}{1}
\setcounter{footnote}{0}
\setcounter{equation}{0}
\noindent

\renewcommand{\thefootnote}{\arabic{footnote}}
\vspace*{-1.5cm}

\setcounter{footnote}{0}

\section{Introduction}
\setcounter{equation}{0}
\subsection{Status of the charm quark mass calculations}
During the last few years a great effort has been 
devoted to the precise determination of the quark masses. 
Recent reviews about the present situation 
concerning the computation of the light quark masses  
can be found in refs.~\cite{vittorio,leutwyler}. 
As for the heavies, most of the studies 
done so far were focused on determining the $b$-quark mass. 
The charmed quark escapes the precision computation mainly 
because it is too heavy for the chiral 
perturbation theory to apply, and yet too light for the heavy 
quark expansion to set in.

A complete account of the presently available estimates of 
the charm quark mass value is given in the PDG review~\cite{Groom:2000in}. 
They estimate the 
charm quark mass to be in  the range~\footnote{Ref.~\cite{Groom:2000in} 
also contains a complete list of references.}
\bea
1.15 \ \gev \ \leq m_c^\msbar(m_c)\ \leq \ 1.35\ \gev \ .
\eea
Very recently, two new QCD sum rule computations of this quantity 
appeared~\cite{jamin,penarrocha}.  
After improving the calculation of the moment QCD sum rules 
for the charmonium states, the new result of ref.~\cite{jamin} is 
$m_c^\msbar(m_c)\ = \ 1.23(9)$~GeV. Adopting quite a different 
QCD sum rule technique, in ref.~\cite{penarrocha} $m_c^\msbar(m_c)\ = \ 
1.37(9)$~GeV was obtained.

On the side of the lattice QCD computations there were a few attempts 
to compute the charm quark. 
\begin{itemize}
\item[--]By combining the QCD sum rule methodology with the lattice QCD computation 
of the moments of the heavy-heavy vector current correlation function, the authors of 
ref.~\cite{bochkarev} obtained $m_c^\msbar(m_c) = 1.22(5)$~GeV.
More complete discussion of the systematic uncertainties involved in their
computation has been made in ref.~\cite{philippe}. 
Unfortunately, this (elegant) method has not been followed by the other 
lattice groups.
\item[--]
From the lattice QCD with Wilson quarks and by using the Ward identities, in 
ref.~\cite{giusti} a much larger value has been obtained, namely 
$m_c^\msbar(m_c) = 1.71(3)(20)$~GeV. 
If we use their recent re-estimate of the mass renormalization constant~\cite{leo}, 
the above value gets down to $m_c^\msbar(m_c) = 1.50(3)(18)$~GeV.
\item[--]
Finally, in ref.~\cite{fermilab} the charm quark mass has been computed by adopting 
the Fermilab way of interpreting the lattice QCD beyond the lattice cut-off. The quoted 
result is $m_c^\msbar(m_c)= 1.33(8)$~GeV.
\end{itemize}
Since the actual situation with the lattice results is not clear, 
we decided to take advantage of the presently available non-perturbatively 
determined parameters, which are necessary for the complete elimination  of the 
discretization effects that are linear in the lattice spacing ({\it i.e.} of ${\cal O}(a)$),
and to make an estimate of the charm quark mass from the data that we obtained by 
working with the (relatively) fine grained lattice ($a \approx 0.07$~fm). 
Our result is
\bea
m_c^\msbar(m_c)  = 1.26(3)(12)~\gev \ .
\eea

\subsection{Computation of the charm quark mass on the lattice}

To compute the charm quark mass, we rely on the standard two strategies:
\begin{itemize}
\item {\sf{Vector Ward identity (VWI)}}
\bea
\label{VWI}
m_Q^{\rm (VWI)}(\mu) = Z_m(\mu) m_Q^{\rm (VWI)} = 
Z_m(\mu) {1 \over 2} \left( {1\over \kappa_Q} - {1\over
\kappa_{cr} }\right) \ ,
\eea
where the critical value of the Wilson hopping parameter ($\kappa_{cr}$) corresponds to the
chiral limit.~\footnote{Unless the physical units are explicitely displayed, 
all the quark masses mentioned in this section are assumed to be in lattice units.} 
Since the lattice cutoff is finite and the charm quark mass is not negligible, 
it is highly important to improve the renormalization constant out of the chiral limit:
\bea
Z_m(\mu) = {1\over Z_S^{(0)}(\mu)} ( 1 + b_m m_Q^{\rm (VWI)})\;,
\eea
where the value of $b_m=-b_S/2$, has been determined non-perturbatively in 
ref.~\cite{sint}, and $Z^{(0)}_S(\mu)$ is the renormalization constant of 
the scalar density computed in the chiral limit, which will be given in the next section.
\item  {\sf Axial-vector Ward identity (AWI)} 
\bea
\label{AWI}
m_Q^{\rm (AWI)}(\mu) + m_q^{\rm (AWI)}(\mu)=  {\overline Z_m(\mu)}  
{ \langle \displaystyle{\sum_{\vec x}} \partial_4 A_4^{\rm I}
(x) P^\dagger(0)\rangle \over \langle\displaystyle{\sum_{\vec x}} P (x) P^\dagger(0)\rangle }  \ ,
\eea
where $A_\mu = \bar Q \gamma_\mu \gamma_5 q$ and $P = \bar Q  \gamma_5 q$ are the
axial vector current and the pseudoscalar density, respectively. 
For the full elimination of ${\cal O}(a)$
effects, the bare lattice axial
current is improved in the chiral limit as $A_\mu^{\rm I} = A_\mu + c_A \partial_\mu P$, 
with $c_A$ already determined non-perturbatively~\cite{lanl,luscher}. We used the symmetric 
definition of the derivative, {\it e.g.} $\partial_4 f = (f(t+ 1)-f(t- 1))/2$. 
The mass renormalization 
constant is equal to $\overline Z_m(\mu) = Z_A/Z_P(\mu)$ and it is improved as
\bea
{\overline Z_m(\mu)} = {Z^{(0)}_A\over Z^{(0)}_P(\mu)} \biggl(1 + (b_A - b_P) 
{m_Q^{\rm (VWI)} + m_q^{\rm (VWI)}\over 2} \biggr)\;.
\eea
Again, the value of $(b_A - b_P)$ has been computed
non-perturbatively~\cite{sint,lanl}, whereas the value of $\overline
Z_m^{(0)}(\mu)=Z^{(0)}_A/Z^{(0)}_P(\mu)$,
 will be provided in the next section.
\end{itemize}

\section{Lattice details and results}

\subsection{A few details on the lattice computation}
We work with two sets of lattice data, generated at $\beta = 6.2$ by using 
the non-perturbatively improved Wilson action ($c_{SW}=1.614$~\cite{luscher}). Each set 
contains 200 independent $SU(3)$ gauge field configurations.
The value of the critical parameter, $\kappa_{cr}$, is fixed by requiring the 
bare $m_q^{\rm (AWI)}\to 0$. 
\begin{table}
\vspace*{-5mm}
\begin{center}
\begin{tabular}{|c||c|c|c||c|c|c|}
  \hline
\hspace{-5.mm}{\phantom{\huge{l}}}\raisebox{-.2cm}{\phantom{\Huge{j}}}
Lattice ($\beta = 6.2$) & \multicolumn{3}{c||}{\underline{$24^3 \times 64 $} ($\kappa_{cr}=0.13583(5)$)}&
\multicolumn{3}{c|}{\underline{$24^3\times 48$} ($\kappa_{cr}=0.13577(3)$)} \\
 \hline
{\phantom{\Large{l}}}\raisebox{.2cm}{\phantom{\Large{j}}}
$\kappa_1$--$\kappa_2$ & 
$m^{\rm (AWI)}$ & $m^{\rm (VWI)}$& $M_P$& $m^{\rm (AWI)}$ & $m^{\rm (VWI)}$& $M_P$ \\
\hline
{\phantom{\Large{l}}}\raisebox{.2cm}{\phantom{\Large{j}}}\hspace{-4.5mm}
$0.1344$--$0.1344$ & 0.0393(15) & 0.0391(13) &0.305(2) 
                   & 0.0399(14) & 0.0376(9)  & 0.307(2)\\
{\phantom{\Large{l}}}\raisebox{.2cm}{\phantom{\Large{j}}}\hspace{-4.5mm}
$0.1349$--$0.1349$ & 0.0249(10) & 0.0253(13) &0.244(2) 
                   & 0.0241(12) & 0.0237(9) & 0.245(2) \\
{\phantom{\Large{l}}}\raisebox{.2cm}{\phantom{\Large{j}}}\hspace{-4.5mm}
$0.1352$--$0.1352$ & 0.0162(9) & 0.0171(13) &0.200(3) 
                   & 0.0155(11)& 0.0154(9) & 0.200(2)  \\  \hline \hline
{\phantom{\Large{l}}}\raisebox{.2cm}{\phantom{\Large{j}}}\hspace{-4.5mm}
$0.125$--$0.1344$ & 0.176(2) & 0.179(1) &0.690(1) 
                   & 0.172(2) & 0.177(1)  & 0.693(2)\\
{\phantom{\Large{l}}}\raisebox{.2cm}{\phantom{\Large{j}}}\hspace{-4.5mm}
$0.125$--$0.1349$ & 0.167(2) & 0.172(1) &0.672(2) 
                   & 0.162(2) & 0.170(1) & 0.675(2) \\
{\phantom{\Large{l}}}\raisebox{.2cm}{\phantom{\Large{j}}}\hspace{-4.5mm}
$0.125$--$0.1352$ & 0.163(3) & 0.168(1) &0.661(2) 
                   & 0.158(3)& 0.166(1) & 0.663(2)  \\  \hline 
{\phantom{\Large{l}}}\raisebox{.2cm}{\phantom{\Large{j}}}\hspace{-4.5mm}
$0.122$--$0.1344$ & 0.221(3) & 0.228(1) &0.786(1) 
                   & 0.217(3) & 0.226(1)  & 0.789(2)\\
{\phantom{\Large{l}}}\raisebox{.2cm}{\phantom{\Large{j}}}\hspace{-4.5mm}
$0.122$--$0.1349$ & 0.212(3) & 0.221(1) &0.768(2) 
                   & 0.209(3) & 0.220(1) & 0.771(2) \\
{\phantom{\Large{l}}}\raisebox{.2cm}{\phantom{\Large{j}}}\hspace{-4.5mm}
$0.122$--$0.1352$ & 0.207(3) & 0.217(1) &0.757(2) 
                   & 0.202(3)& 0.215(1) & 0.761(3)  \\  \hline 
{\phantom{\Large{l}}}\raisebox{.2cm}{\phantom{\Large{j}}}\hspace{-4.5mm}
$0.119$--$0.1344$ & 0.267(3) & 0.280(1) &0.876(1) 
                   & 0.261(3) & 0.278(1)  & 0.878(2)\\
{\phantom{\Large{l}}}\raisebox{.2cm}{\phantom{\Large{j}}}\hspace{-4.5mm}
$0.119$--$0.1349$ & 0.258(3) & 0.273(1) &0.858(2) 
                   & 0.252(3) & 0.271(1) & 0.861(3) \\
{\phantom{\Large{l}}}\raisebox{.2cm}{\phantom{\Large{j}}}\hspace{-4.5mm}
$0.119$--$0.1352$ & 0.251(3) & 0.269(1) &0.847(3) 
                   & 0.248(4)& 0.267(1) & 0.851(4)  \\  \hline 
		   \end{tabular}
\caption{\label{tab1}{\small \sl Bare quark masses obtained by using 
the vector and the axial Ward identity on the lattice (see eq.~(\ref{defM})). 
The results are given in lattice units. On the subset of the configurations
gathered on the lattice $24^3\times 64$, in ref.~\cite{strano} we computed the
strange quark mass.}}
\end{center}
\vspace*{-5mm}
\end{table}
The values of $\kappa_{cr}$, along with the values of the bare quark masses for the degenerate light quark 
combinations, as well as for the non-degenerate (heavy-light) ones, are given in tab.~\ref{tab1}. 
Each $m^{\rm (AWI)}$ and $m^{\rm (VWI)}$ are obtained as
\bea \label{defM}
m^{\rm (AWI)} &=& {1\over 2} { \langle \displaystyle{\sum_{\vec x}} \partial_4 A_4^{\rm I}
(x) P^\dagger(0)\rangle \over \langle\displaystyle{\sum_{\vec x}} P (x) P^\dagger(0)\rangle }  \ ,\cr
&& \cr
m^{\rm (VWI)} &=& {1\over 4} \left( {1\over \kappa_Q} +{1\over \kappa_q} - {2\over \kappa_{cr}}  \right)\ ,
\eea
where in the improved axial current, $A_4^{\rm I} = \bar Q\gamma_4\gamma_5 q + c_A \partial_4 \bar Q
\gamma_5 q$, we used $c_A=-0.04(1)$~\cite{lanl,luscher}.
The statistical errors quoted in this work are obtained by using the 
standard jackknife procedure (with 5 configurations per jack). The quark masses listed in 
tab.~\ref{tab1} are the bare lattice ones, 
that we now need to renormalize. We next discuss the computation of the mass 
renormalization constants.

\subsection{Mass renormalization constants}

To evaluate the mass renormalization constants in the chiral limit we use
the non-perturba\-ti\-ve method~\cite{guido} which allows one to compute these constants in 
the continuum $\ri$ scheme. The use of this method out 
of the chiral limit, however, requires the improvement of the off-shell quantities 
for which several 
new counterterms appear, each with a coefficient that is to be fixed non-perturbatively 
(for more details, see ref.\cite{massimo}). For that reason, in this paper, 
we will employ the method 
to compute the renormalization constants in the chiral limit only~\cite{light}. 
To improve the renormalization constants out of the chiral limit, 
we will rely on the results of ref.~\cite{sint,lanl} in which the corresponding 
coefficients 
were computed non-perturbatively with the results, $b_m=-0.69(1)$, $b_A-b_P=0.04$, 
at $\beta=6.2$.

\begin{figure}[h]
\begin{center}
\begin{tabular}{@{\hspace{-.95cm}}c}
\epsfxsize18.0cm\epsffile{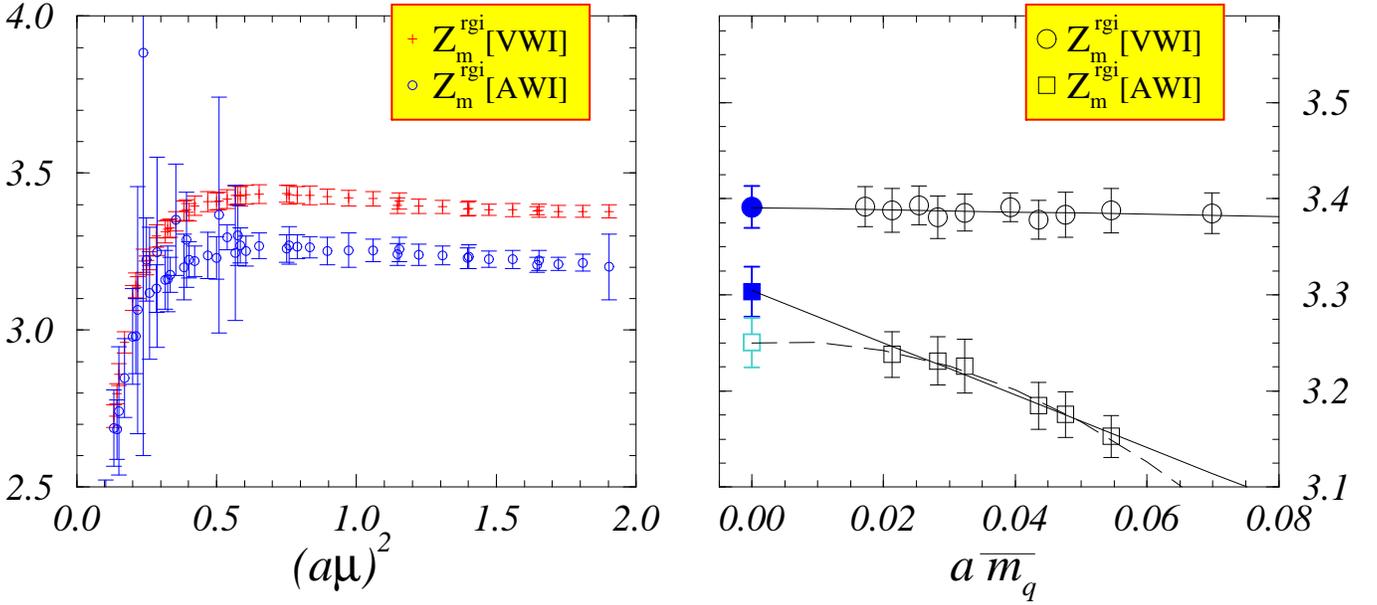}    \\
\end{tabular}
\vspace*{-.85cm}
\caption{\label{fig1}{\small \sl Mass renormalization constant: In the left figure we show the 
renormalization constants divided by the perturbative 4-loop anomalous dimension for the case of
VWI ($Z_m =1/Z_S$) and AWI ($\overline Z_m =Z_A/Z_P$) for a specific combination of $\kappa_{1} = 0.1349$
and $\kappa_{2}=0.1352$. On the right figure we show the extrapolation of the mass renormalization
constants to the chiral limit. Dashed line depicts the quadratic extrapolation to ${\overline
Z}_m^{(0)\rm rgi}$.}}
\end{center}
\end{figure}

We first focus on the mass renormalization constant, $Z_m^{(0)}(\mu) =
1/Z_S^{(0)}(\mu)$. To that end, we combine the quark propagators 
of 4 different values of $\kappa_q$ in 10 different
combinations (degenerate and non-degenerate in light quark mass), compute 
the amputated vertex function $\Gamma_S(\kappa_i, \kappa_j; a\mu )$ and 
impose the standard RI/MOM renormalization condition~\footnote{Besides the 
values of the Wilson hopping parameters corresponding to the light quark 
masses that we already listed in tab~\ref{tab1}, for this computation 
we also use the quark propagator with $\kappa_q=0.1333$.}  
\bea
 { 1 \over Z_S^{(0)}(\mu )}\ =\ \lim_{\kappa_{1,2}\to \kappa_{cr}}  { 1 \over Z_S(\kappa_{1}, \kappa_{2};\mu )}\ =\ \lim_{\kappa_{1,2}\to \kappa_{cr}} \left. {\Gamma_S(\kappa_i, \kappa_j;  p )\over 
 Z_q^{1/2}(\kappa_i;  p )\ Z_q^{1/2}(\kappa_j;  p )} 
\right|_{p^2 = \mu^2} \,, 
\eea
where $Z_q(\mu)$ is the quark field renormalization constant which is easily 
extracted by imposing the vector Ward identity on the quark propagator. 
Such an obtained $Z_m^\ri (\mu) = 1/Z_S^\ri(\mu)$ is then converted to the 
renormalization group 
invariant (RGI) form, $Z_m^{\rm rgi}=Z_m^\ri (\mu)/c^\ri (\mu)$, by using the 
available mass anomalous 
dimension coefficients up to 4-loops encoded in the function $c^\ri (\mu)$, 
computed in the same 
$\ri$ scheme~\cite{retey}.~\footnote{The explicit form of the 
function $c^\ri (\mu)$ is also given in Appendix of the present letter.} 
For consistency, we also use the 4-loop expression for the running 
coupling~\cite{beta4}, 
and set $n_F=0$ with (quenched) $\Lambda^{(n_F=0)}_{\rm QCD} = 0.25$~GeV~\cite{pepe,capitani}. 
A typical
situation is shown in fig.~\ref{fig1}(left). For every mass combination, 
we then extract 
$Z_m^{\rm rgi}(\mu)$ by fitting to a constant on the plateau $1.0 \leq (a \mu)^2 
\leq 1.8$. 
After extrapolating in $a\overline m_q = (m_1^{\rm (VWI)} + m_2^{\rm (VWI)})/2$ 
to the chiral limit, we get 
\bea
Z_m^{(0) \rm rgi} = 3.391(22) \, .
\eea
For an easier comparison with the other determinations of this renormalization 
constant we express this result
also in the $\msbar$ scheme:
\bea
Z_m^{(0) \msbar } (2\ \gev)= 1.332(9) \, .
\eea
As it can be 
observed from fig.~\ref{fig1}(right), the extrapolation to the chiral limit 
in this case is very smooth.

Next, we discuss the computation of the second mass renormalization constant, 
{\it i.e.} the one needed to compute the quark mass by using the axial Ward identity
 ($\overline Z_m^{(0)}(\mu) = Z_A^{(0)}/Z_P^{(0)}(\mu)$). We use the proposal 
 of ref.~\cite{leo} which, by judiciously combining the Ward identities, allows one to 
 alleviate the problem of the contamination by the Goldstone boson~\cite{alain}. The renormalization 
 condition, by which this task is achieved for the mass renormalization constant, can be 
simply written as
\bea
 {Z_A^{(0)}\over Z_P^{(0)}}(\mu)\ =\  \lim_{\kappa_{1,2}\to \kappa_{cr}}  {Z_A\over Z_P}(\kappa_1,\kappa_2; \mu)\ = \ \lim_{\kappa_{1,2}\to \kappa_{cr}}  \left.
{ m_1^{\rm (VWI)}\Gamma_P(\kappa_1; p) - m_2^{\rm (VWI)}\Gamma_P(\kappa_2; p)\over 
(m_1^{\rm (VWI)} - m_2^{\rm (VWI)}) \Gamma_A(\kappa_1,\kappa_2; p)}\right|_{p^2 = \mu^2}.
\eea
Obviously, we can use only the non-degenerate quark mass combinations which, for our 4 
values of $\kappa_q$, means 6 combinations. As in the previous 
case, we convert our result from the $\ri$ to the renormalization group invariant constant 
(at
4-loop level) and fit in the same window as before, $1.0\leq (\mu a)^2 \leq 1.8$ (see 
fig.~\ref{fig1}(left)). With such 
extracted values for ${\overline Z}_m^{\rm rgi}$, for each combination of the Wilson 
hopping parameters ($\kappa_1,\kappa_2$), we then  
extrapolate to the chiral limit. This is also illustrated in fig.~\ref{fig1}(right). 
Contrary to the first case, the mass dependence of the 
${\overline Z}_m^{\rm rgi}$ is more pronounced. We extrapolate to the chiral limit 
linearly (filled square in fig.~\ref{fig1}) to get our central value. In addition, 
we perform the quadratic extrapolation (the result of which is depicted by an empty square in 
fig.~\ref{fig1}), and the difference between this and the central value is incorporated in 
the systematic uncertainty.  
Our result is
\bea \label{Zap}
{\overline  Z}_m^{(0) \rm rgi} = 3.303(26)^{+0.000}_{-0.051} \,,
\eea
which in the $\msbar$ scheme reads
\bea
{\overline  Z}_m^{(0) \msbar } (2\ \gev )= 1.297(10)^{+0.000}_{-0.020} \,.
\eea
This result agrees well with the one of 
ref.~\cite{capitani}, ${\overline  Z}_m^{(0) \msbar } (2\ \gev ) = 1.316(14)(17)$.

\subsection{Putting it all together}

Now we combine all the results from tab.~\ref{tab1} with the renormalization constants 
discussed in the previous section to get the renormalization group invariant 
quark masses.  
At this stage (after including the renormalization constants) one can identify the results of 
the axial Ward identity as the sum of the heavy and the light quark mass. Since we also 
computed the light quark mass separately, we can now simply subtract it
from the sum and work with the heavies only. The results are presented in tab.~\ref{tab2}. In the same table we
give the values of the heavy-light mesons for which the light quark mass has been interpolated 
to the light $s$-quark in a usual way (see {\it e.g.} ref.~\cite{hl1}). The mass of
the $D_s$ meson in lattice units is $M_{D_s} \simeq 0.73(3)$, where we use $a^{-1}(m_{K^*}) =
2.7(1)$~GeV. Thus the charm quark mass is to be found through an interpolation between the 
results for $\kappa_Q = 0.125$  and $\kappa_Q = 0.122$.
\begin{table}[h!!]
\vspace*{-1mm}
\begin{center}
\begin{tabular}{|c||c|c|c||c|c|c|}
  \hline
\hspace{-4.mm}{\phantom{\huge{l}}}\raisebox{-.2cm}{\phantom{\Huge{j}}}
$\diamond$& \multicolumn{3}{c||}{\underline{$24^3 \times 64 $}} &
\multicolumn{3}{c|}{\underline{$24^3\times 48$}} \\
 \hline
{\phantom{\Large{l}}}\raisebox{.2cm}{\phantom{\Large{j}}}
$\kappa_Q$ & 
$m_Q^{\rm rgi}$ [AWI] & $m_Q^{\rm rgi}$ [VWI]& $M_{P_s}$& 
$m_Q^{\rm rgi}$ [AWI] & $m_Q^{\rm rgi}$ [VWI]& $M_{P_s}$ \\
\hline
{\phantom{\Large{l}}}\raisebox{.2cm}{\phantom{\Large{j}}}\hspace{-4.5mm}
$0.125$ & 1.032(13) & 0.868(3) & 0.676(3) 
                   & 1.000(14) & 0.864(2)  & 0.681(5)\\
{\phantom{\Large{l}}}\raisebox{.2cm}{\phantom{\Large{j}}}\hspace{-4.5mm}
$0.122$ & 1.334(16) & 1.013(2) &0.771(3) 
                   & 1.306(16) & 1.011(1) & 0.777(4) \\
{\phantom{\Large{l}}}\raisebox{.2cm}{\phantom{\Large{j}}}\hspace{-4.5mm}
$0.119$ & 1.638(20) & 1.141(1) &0.861(3) 
                   & 1.592(19)& 1.139(1) & 0.866(4)  \\  \hline 
		   \end{tabular}
\caption{\label{tab2}{\small \sl Renormalized heavy quark masses directly accessed from our
lattice. }}
\end{center}
\vspace*{-5mm}
\end{table}
Notice that in tab.~\ref{tab2} we give also the results for a larger quark
mass corresponding to $\kappa_Q=0.119$. Although this value is not necessary 
for our final result for the charm quark mass, it will be helpful in assessing the
amount of the systematic uncertainties which will be discussed in the next subsection.

Now, to get the value of the charm quark mass, we need to interpolate in the heavy 
meson mass to $M_{D_s}$ and then simply read off the charm quark mass. To do so we 
need to choose
an interpolating formula. We consider the following ones:
\begin{itemize}
\item[(i)] $M_{P_s} = a_0 + a_1 m_Q + a_2 m_Q^2$;
\item[(ii)] $M_{P_s} = b_0 + b_1/m_Q + b_2/m_Q^2$;
\item[(iii)] $M_{P_s} = c_0 + c_1 m_Q + c_2/m_Q$.
\end{itemize}
The first (i) is the naive linear interpolation ($a_2=0$), the second (ii) comes from the heavy 
quark expansion and the third (iii) is the hybrid of the two. For (iii) we obviously need 
at least three points, and thus we have to use also the heaviest of our quarks from
tab.~\ref{tab2}. The complete situation is presented in tab.~\ref{tab3}. 
Since the heavier quark is more prone to the ${\cal O}\left((a m)^2\right)$ 
artifacts, we first concentrate our discussion on the results of the first two 
interpolations in which we set $a_2=b_2=0$. 
\begin{table}[h!!]
\begin{center}
\begin{tabular}{|c||c|c||c|c|}
  \hline
\hspace{-4.mm}{\phantom{\huge{l}}}\raisebox{-.2cm}{\phantom{\Huge{j}}}
 & \multicolumn{2}{c||}{\underline{$24^3 \times 64 $} (Lattice-I)} &
\multicolumn{2}{c|}{\underline{$24^3\times 48$} (Lattice-II)}\\
 \hline
{\phantom{\Large{l}}}\raisebox{.2cm}{\phantom{\Large{j}}}
$\kappa_Q$ & 
$m_c^{\rm rgi}$ [AWI] & $m_c^{\rm rgi}$ [VWI] &
$m_c^{\rm rgi}$ [AWI] & $m_c^{\rm rgi}$ [VWI] \\
\hline
{\phantom{\Large{l}}}\raisebox{.2cm}{\phantom{\Large{j}}}\hspace{-4.5mm}
Form (i) ($a_2=0$)& 1.190(84) & 0.944(39)  
                   & 1.136(100) & 0.929(47)  \\
{\phantom{\Large{l}}}\raisebox{.2cm}{\phantom{\Large{j}}}\hspace{-4.5mm}
Form (ii)($b_2=0$) & 1.171(83) & 0.939(39)  
                   & 1.126(98) & 0.923(46)  \\  \hline
{\phantom{\Large{l}}}\raisebox{.2cm}{\phantom{\Large{j}}}\hspace{-4.5mm}
Form (i) & 1.186(84) & 0.945(39)  
                   & 1.135(100) & 0.930(47)  \\
{\phantom{\Large{l}}}\raisebox{.2cm}{\phantom{\Large{j}}}\hspace{-4.5mm}
Form (ii) & 1.191(82) & 0.947(39)  
                   & 1.143(98) & 0.932(46)  \\ 
{\phantom{\Large{l}}}\raisebox{.2cm}{\phantom{\Large{j}}}\hspace{-4.5mm}
Form (iii) & 1.187(84) & 0.954(39)  
                   & 1.137(100) & 0.931(47)  \\  \hline 
		   \end{tabular}
\caption{\label{tab3}{\small \sl Charm quark mass in lattice units as obtained by using the 
interpolating formulae (i), (ii)  and (iii), as discussed in the text. The first two are obtained 
without using the heaviest of our quarks (the one corresponding to $\kappa_Q=0.119$).}}
\end{center}
\vspace*{-5mm}
\end{table}
The results of the two interpolation formulae are totally consistent with each other, and we
choose to quote the first one as our central number.

To get the final result, that can be confronted to the results of other approaches, we need to convert 
our values to the $\msbar$ scheme and express it in the physical units by using $a^{-1}(m_{K^\ast})=2.7(1)$~GeV. 
After recalling that $m_c^{\rm rgi} \cdot c^\msbar(\mu) =  m_c^\msbar(\mu)$, 
where the function $c^\msbar(\mu)$ is
known to 4-loop accuracy~\cite{vermas} (see App. of the present letter), 
one can easily 
solve that equation to obtain the standard value $m_c^\msbar(m_c)$. Our results are:
\bea 
{\rm Lattice-I}&&m_c^\msbar(m_c)_{\rm VWI} = 1.144(3)~\gev \;,\nonumber\\
&&m_c^\msbar(m_c)_{\rm AWI}  = 1.373(34)~\gev \label{eq9}\;,\\
&& \nonumber\\
{\rm Lattice-II}&&m_c^\msbar(m_c)_{\rm VWI} = 1.132(3)~\gev \;,\nonumber\\
&&m_c^\msbar(m_c)_{\rm AWI}  = 1.325(43)~\gev \;.\label{eq10}
\eea
The above results are obtained by using $n_F=0$ and $\Lambda^{n_F=0}_{\rm QCD}=0.25$~GeV. 
We see that for both sets of our lattice data the two equivalent
methods (VWI and AWI) give different results. The reason for that discrepancy most probably comes from
the lattice artifacts which are $\propto (a m_Q)^n$ ($n\geq 2$). One way of seeing that  
is to include the higher order effects at tree-level by employing the so-called EKLM factors~\cite{eklm} 
(see also the discussion in~\cite{tassos}). The leading effect 
of the EKLM factors to our result is $\propto (a m_Q)^2$ and it modifies the RGI 
charm quark mass as follows
\bea
&& m_c^{\rm rgi}{\rm [VWI]} \longrightarrow \left[ 1 +{(m_c^{\rm (VWI)})^2\over 12} \right]
m_c^{\rm rgi}{\rm [VWI]} \;, \nonumber\\
&&\nonumber\\
&& m_c^{\rm rgi}{\rm [AWI]} \longrightarrow   \left[ 1 -{(m_c^{\rm (VWI)})^2\over 6} \right] m_c^{\rm rgi}{\rm [AWI]}\,, 
\eea
where $m_c^{\rm (VWI)}$ stands for the bare lattice charm quark mass whose value is
$m_c^{\rm (VWI)}=0.375(25)$. 
When the above modification is included and we pass onto $m_c^\msbar(m_c)$, 
from our Lattice-I simulation, we get
\bea
&&m_c^\msbar(m_c)_{\rm VWI} = 1.161(3)~\gev \,,\nonumber\\
&&m_c^\msbar(m_c)_{\rm AWI}  = 1.342(33)~\gev \,.
\eea
After comparing these to the results~(\ref{eq9}), we see that the inclusion 
of the tree level ${\cal O}\left((a m_Q)^2\right)$ effects makes our two 
results getting closer to each other, although it is not sufficient to remove the bulk of 
${\cal O}\left(a^2\right)$ corrections.

\subsection{Systematic uncertainties and our final result}
We will now briefly summarize the sources of systematic uncertainties and comment each one
of them.
\begin{itemize}
\item The most important source of the systematic error 
are the lattice artifacts of ${\cal O}((a m)^n)$ $(n\geq 2)$. 
It is therefore highly important to repeat our calculation on the lattice 
with a smaller lattice spacing. 
As our central result we will quote the average of the two methods (VWI and AWI), 
for both our sets of data as given in eqs.~(\ref{eq9},\ref{eq10}). 
The larger statistical error will be attributed to our final result while the difference
between any of the two methods and the averaged one is included in the systematic uncertainty. 
In other words we have:
\bea
{\rm Lattice-I}&&m_c^\msbar(m_c)  = 1.26(3)(11)~\gev \ ,\nonumber\\
&& \nonumber\\
{\rm Lattice-II}&&m_c^\msbar(m_c)  = 1.23(4)(10)~\gev\,. 
\eea
The results of our two simulations are very close to each other 
and for our final estimate of the charm quark mass we will quote the 
value obtained from the first lattice, which has the larger temporal extension. 

\item In our analysis, we used the value of the lattice spacing 
$a^{-1}=2.71(11)$~GeV, as obtained from the comparison of the physical $m_{K^\ast}$ 
and the one we computed on the lattice. 
We checked that if, instead of the above value, we
use $a^{-1}=2.83(15)$ as obtained from the $f_K$ decay constant, 
the final value of the charm quark gets larger by $2\%$.

\item To be conservative, when taking the average of the masses obtained 
by using the two equivalent 
methods, we quoted the larger statistical error. However, the final result does not contain 
$-1.5\%$ of the error on the renormalization constant $\overline Z^{\rm rgi}_m$ which we 
discussed in the text (see eq.~(\ref{Zap})). 
Although the value of the renormalization constant $Z^{\rm rgi}_m$ does not suffer from the
same uncertainty, to be on the safe side, we will add $-1.5\%$ of error to our final result.

\item Whenever needed,  we used 
$\Lambda_{\rm QCD}^{n_F=0}=0.25$~GeV. Varying this quantity by $10\%$ allows one to cover 
all the presently available lattice estimates for  $\Lambda_{\rm QCD}^{n_F=0}$~\cite{pepe,capitani}. 
The impact of that variation on the final charm quark mass value is $\pm 1.6\%$.

\item In ref.~\cite{lanl}, the authors also estimate the discretization errors on the
improvement coefficient $b_A - b_P$. We have varied this coefficient by $\pm 0.03$ which
introduces the change in the central value for the charm quark mass by less that $1\%$ 
(more precisely by $\pm 0.6\%$).

\item As mentioned in the previous subsection, the result obtained from the naive 
(linear) interpolation practically coincides with the one that we get when employing 
the interpolation motivated by the heavy quark expansion. Since we have the third (heavier) 
quark mass, we checked that the quadratic terms to both formulae~(i,ii), or the use of the 
{\sl ``hybrid''} formula~(iii) affects the value of the charm quark mass by no more 
than $1\%$ (see tab.~\ref{tab3}).

\item For determination of the charm quark mass we used the masses of the heavy-light 
pseudoscalar mesons with the light quark mass (linearly) interpolated to the strange one. 
We checked that the quadratic interpolation in the light quark mass to get 
 $M_{P_s}$, does not make any influence on our final charm quark mass value.
In addition, we verified that our results remain unchanged if we use the vector 
heavy-light mesons instead of the pseudoscalars. 
\end{itemize}
We sum the above sources of errors quadratically and obtain
\bea
m_c^\msbar(m_c)  = 1.26(3)(12)~\gev \ ,
\eea
which is our final result. We note also that in  passing from the RGI to the
$\msbar$ value, we might have as well used $n_F=4$ and the value 
$\alpha_s(m_\tau)=0.334(22)$~\cite{aleph}. This, however, does not modify the above
result, namely we obtain $m_c^\msbar(m_c)  = 1.27(3)(11)~\gev$.

\section{Summary and Perspectives}

In this letter we have computed the charm quark mass on the lattice by taking the advantage 
of the full improvement of the Wilson action and operators by which all the lattice artifacts 
linear in lattice spacing are absent. Therefore, the computation of the charm quark mass on 
a reasonably fine grained lattice is expected to lead to the result close to the 
continuum limit. 
Results of our simulations, obtained at single value of the lattice spacing, indicate that 
the lattice artifacts ${\cal O}(a^2)$ are sensible.
As a consequence the two equivalent methods to compute the quark mass on the lattice yield 
different values. This fact largely dominates the systematic uncertainty of our calculation, 
and for the precise determination of the charm quark mass it is therefore important 
to go to ever smaller lattice spacings. 
Finner lattices, in turn, require larger lattices (to keep the same physical volume) 
and thus more powerful computing resources. 
By using APE1000 we plan to make such a study in the near future. 
We stress also that our computation has been performed in the quenched approximation. 
The (unknown) uncertainty introduced by quenching is not included in our systematic error
estimate. A na\"{\i}ve guess points towards a decrease of the quenched value by 
$\sim 5\%$.~\footnote{ 
This ``guesstimate'' is based on the present observations according to which the 
unquenching of the light quark masses reduces their values by $\sim 10\%$, 
whereas $m_b$ gets smaller by less than $2\%$~\cite{vittorio}. }
However we prefer to quote our result as the quenched one and wait for the partially 
unquenched computations to assess the amount of this source of systematic errors.

\section*{Appendix}
The RGI quark mass is defined as
\bea
m^{\rm rgi} ={\ m^\msbar(\mu)\ \over c^\msbar(\mu)} = {\ m^\ri(\mu)\ \over c^\ri(\mu)} \;,
\eea
where the functions $c^\msbar(\mu)$ and  $c^\ri(\mu)$ are known to 4-loop accuracy~\cite{retey,vermas}.
For $n_F=0$, they read
\bea
&&\hspace*{-14mm}c^\msbar (\mu) = a_s(\mu)^{4/11} \biggl( 1\ +\ 0.68733\ a_s(\mu)\ +\ 1.51211\ a_s(\mu)^2\ +\ 
4.05787 \ a_s(\mu)^3\biggr),\\
&& \nonumber \\
&&\hspace*{-14mm}c^\ri (\mu) = a_s(\mu)^{4/11} \biggl( 1\  +\  2.02067\  a_s(\mu)\  +\  14.21925\ 
a_s(\mu)^2\ +\  138.30689\ a_s(\mu)^3\biggr),
\eea
where, for short, we write $a_s(\mu) \equiv \alpha_s(\mu)/\pi$.

\end{document}